\documentclass[namedreferences]{solarphysics}

\usepackage[hyperref,optionalrh,showbiblabels]{spr-sola-addons} 
\usepackage{graphicx}        
\usepackage{color}           
\ifx \doiurl    \undefined \def \doiurl#1{\href{http://dx.doi.org/#1}{\textsf{DOI}}}\fi
\ifx \adsurl    \undefined \def \adsurl#1{\href{http://adsabs.harvard.edu/abs/#1}{\textsf{ADS}}}\fi
\ifx \arxivurl  \undefined \def \arxivurl#1{\href{http://arxiv.org/abs/#1}{\textsf{arXiv}}}\fi

\newcommand{\etal}{{\it et al.}}

\newcommand{\adv}{    {\it Adv. Space Res.}}

\newcommand{\aap}{    {\it Astron. Astrophys.}}

\newcommand{\apjl}{   {\it Astrophys. J. Lett.}}

\newcommand{\grl}{    {\it Geophys. Res. Lett.}}

\newcommand{\jgr}{    {\it J. Geophys. Res.}}
\newcommand{\mnras}{  {\it Mon. Not. Roy. Astron. Soc.}}
\newcommand{\nat}{    {\it Nature}}

\newcommand{\solphys}{{\it Solar Phys.}}

\chardef\us=`\_

\begin{document}

\begin{article}
\begin{opening}

\title{A Long-Term Decrease of the Mid-Size Segmentation Lengths Observed in the He~\small{II} \large{(30.4~nm) Solar EUV Emission}}
\author[addressref={aff1},corref,email={leonid@usc.edu}]{\inits{L.}\fnm{Leonid}~\lnm{Didkovsky}}
\author[addressref={aff1},email={wieman@usc.edu}]{\inits{S.}\fnm{Seth}~\lnm{Wieman}}
\author[addressref={aff2},email={ekorogodina@mail.ru}]{\inits{E.V.}\fnm{Elena}~\lnm{Korogodina}}
%
%

\address[id=aff1]{University of Southern California, Space Sciences Center, 835 Bloom Walk, Los Angeles, CA 90089-1341, USA}
\address[id=aff2]{Moscow State Technical University, 5 2nd Baumanskaya, Moscow 105005, Russia}

\runningauthor{L. Didkovsky \textit{et al.}}
\runningtitle{A long-term Decrease of EUV Segmentation Lengths}

\begin{abstract}
Power spectra of segmentation-cell length (a dominant length scale of EUV emission in the transition region) from full-disk He\,{\sc ii} extreme ultraviolet (EUV) images observed by the \textit{Extreme ultraviolet Imaging Telescope} (EIT) onboard the \textit{Solar and Heliospheric Observatory} (SOHO) and the \textit{Atmospheric Imaging Assembly} (AIA) onboard the \textit{Solar Dynamics Observatory} (SDO) during periods of quiet Sun conditions for a time interval from 1996 to 2015 were analyzed. The spatial power as a function of the spatial frequency from about 0.04 to 0.27 (EIT) or up to 0.48 (AIA)~Mm$^{-1}$ depends on the distribution of the observed segmentation-cell dimensions\,--\,a structure of the solar EUV network. The temporal variations of the spatial power reported by
Didkovsky and Gurman (\textit{Solar Phys.} \textbf{289}, 153, \citeyear{Didkovsky14a})
were suggested as decreases at the mid-spatial frequencies for the compared spectra when the power curves at the highest spatial frequencies of 0.5~pix$^{-1}$ were adjusted to match each other. That approach has been extended in this work to compare spectral ratios at high spatial frequencies expressed in the solar spatial frequency units of Mm$^{-1}$. A model of EIT and AIA spatial responses allowed us to directly compare spatial spectral ratios at high spatial frequencies for five years of joint operation of EIT and AIA, from 2010 to 2015. Based on this approach we represent these ratio changes as a long-term network transformation which may be interpreted as a continuous dissipation of mid-size network structures to the smaller-size structures in the transition region. In contrast to expected cycling of the segmentation-cell dimension structures and associated spatial power in the spectra with the solar cycle, the spectra demonstrate a significant and steady change of the EUV network. The temporal trend across these structural spectra is not critically sensitive to any long-term instrumental changes, \textit{e.g.} degradation of sensitivity, but to the change of the segmentation-cell dimensions of the EUV network structure.
\end{abstract}
\end{opening}
\section{Introduction}
     \label{S-Introduction}
 Radiative forcing of the Earth's atmosphere plays a significant role in its thermal and chemical balance \citep{Haigh94, Haigh10}. Effects of heating and cooling are  influenced by long-term solar-cycle  changes. One example of such change compiled from sources that show sensitivity to the changes of solar activity \citep{Hoyt98} is the Maunder Minimum of 1645 to about 1715 \citep{Maunder90}. These observations demonstrate the effects of solar-activity changes during the Maunder Minimum for which low to near-zero sunspot numbers persisted for about six solar cycles (SC) with a SC-averaged period (for SC 1 to 22) of 11 years \citep{Hathaway10}.  It is unclear whether the Mini Ice Age starting about 1650 and associated with a rapid expansion of mountain glaciers was caused by the Maunder Minimum, because the two other known chilly periods around 1770 and 1850 were not directly associated with similar changes of solar activity, \textit{e.g.} with decreased sunspot numbers. From the point of view of finding a correlation between the chilly periods and decreased solar activity, a proxy based on sunspot number is not accurate for the periods of low solar activity, i.e. during solar minima. Another proxy of solar activity, the Total Solar Irradiance (TSI) demonstrates quite small changes (SC maximum to minimum), about 0.15\,\% \citep{Frohlich09}, and it is affected by the calibration uncertainties for instruments on different spacecrafts. The systematic offsets between such measurements complicate accurate determinations of the long-term variations in the absolute TSI, which could detect unusually deep solar minima.
Our analyses of absolute solar EUV irradiance during the prolonged SC 23/24 minimum in 2008\,--\,09 \citep{Didkovsky09} showed a significant decrease of the spectral irradiance in the band around the strong He\,\textsc{ii} emission line of 30.4~nm compared to the level detected for the previous minimum of SC 22/23 in 1996. This decrease was consistent with the unusually low thermosphere neutral density \citep{Solomon10} and the decrease of the Total Electron Content in its sectorial harmonics \citep{Didkovsky14b}. However, some other sources of the anomalously low thermospheric density, such as increased anthropogenic factors, may add a significant amount to the solar EUV forcing \citep{Solomon11}. Some authors \citep{Lean11a, Lean11b} suspect that the detected decrease (about 12\,\%) of the EUV 30.4~nm irradiance \citep{Wieman14} during the latest minimum was produced by the uncorrected degradation of the \textit{Solar Extreme ultraviolet Monitor} (SEM) instrument \citep{Judge98}, a channel of the \textit{Charge Element and Isotope Analysis System} (CELIAS) \citep{Hovestadt95} onboard of \textit{Solar and Heliospheric Observatory} (SOHO). This suggestion is not confirmed by a number of sounding-rocket calibration underflights and comparisons with the data from the \textit{Extreme ultraviolet Spectrophotometer} (ESP) \citep{Didkovsky12}, a channel of the \textit{Extreme ultraviolet Variability Experiment} (EVE) \citep{Woods12} onboard the \textit{Solar Dynamics Observatory} (SDO) \citep{Pesnell12}. ESP, an advanced version of the SEM, uses daily and weekly calibrations to determine channel degradation and, thus, it provides very accurate information about thin-film filters and electronics degradation.

The question of whether or not the SC 23/24 minimum demonstrated a substantial decrease of the EUV irradiance is related to the accuracy of observations at one AU. Another question, more important from the point of view of understanding the causes of such change and possible prediction of its consequences, is whether some solar (internal) sources of this change can be detected and estimated using different methods not sensitive to degradation.

\inlinecite{McIntosh11a,McIntosh11b} studied changes in the solar internal structures associated with this prolonged minimum using Mount Wilson image records in the Ca\,\textsc{ii} K line and in the EUV He\,\textsc{ii} line. The ``watershed segmentation technique'' used by \inlinecite{McIntosh11a} for this study was based on a transformation of the segmentation-cell EUV network (a dominant length scale of EUV emission in the transition region) observed by the \textit{Extreme ultraviolet Imaging Telescope} (EIT) \citep{Delaboudiniere96} onboard the SOHO to analyze the evolution of emission length scale during periods of low solar activity. \inlinecite{Didkovsky14a} used He\,\textsc{ii} full-disk images from EIT and the \textit{Atmospheric Imaging Assembly} (AIA) \citep{Lemen12} onboard the SDO to analyze the spatial power spectra which demonstrated some long-term evolution in the EUV network length scale.
\inlinecite{Williams11}
analyzed Doppler data from the \textit{Michelson Doppler Imager} (MDI) \citep{Scherrer91} onboard the SOHO and found a decrease of supergranular sizes in the photosphere from (35.9\,$\pm$ 0.3~Mm) during the SC minimum in 1996 to (35.0\, $\pm$ 0.3~Mm) during the SC minimum in 2008.

The goal of this study is to investigate the temporal profile of this evolution in the segmentation-cell EUV network length scale for a much longer time interval than from 1996 to 2011 reported by
\inlinecite{Didkovsky14a}.
In part, this may determine whether the profile of the change follows the SC periodicity or demonstrate its independency from the SC periodical behavior.

\section{Data Observations}
     \label{S-Data Observations}

\inlinecite{Didkovsky14a}
showed a change of the EUV network length scale for the 2008\,--\,2010 SC 23/24 minimum, which contained fewer features with a spatial size of three to ten EIT CCD pixels (5 to 18~Mm) than during the 22/23 minimum of 1996. For that comparison we used some days with  low solar activity during 1996, 2008, 2010, and 2011; see Table 1 in
\inlinecite{Didkovsky14a}.

For our analysis in this work we included additional low solar-activity images, EIT for 2012\,--\,2015 and AIA for 2010\,--\,2015, which have been compared to the images from 1996, 2008, and 2010. Thus, the whole time interval, more than 19 years, spans more than 1.5 SC, which allowed us to estimate whether detected changes of the segmentation-cell structure follow the SC periodicity or show a different behavior. An initial procedure used for selecting these low-solar-activity images for the period of more active Sun than the quieter 2008\,--\,10 period was based on two filters. The first filter was to exclude images with solar flares. The second filter was to analyze power spectra of segmentation-cell lengths and remove images that produce noisy power spectra with contamination of the spectral density by some power fluctuations related to large changes of solar EUV irradiance. The full data set used for this analysis is shown in Table 1.

\begin{table}
\caption{EIT and AIA data files grouped for the same year. EIT \textsf{efz} file names include UT time (after a dot). }
\begin{tabular}{cccc} 
\hline
  EIT Date & \textsf{efz} File Name  & AIA Date & Time  \\
\hline
27\,Apr\,96 & \textsf{19960427.033225} &      &     \\
27\,Apr\,96 & \textsf{19960427.092959} &      &     \\
28\,Apr\,96 & \textsf{19960428.003126} &      &     \\
28\,Apr\,96 & \textsf{19960428.062000} &      &     \\
28\,Apr\,96 & \textsf{19960428.175716} &      &     \\
\hline
27\,Nov\,08 & \textsf{20081127.011936} &      &      \\
27\,Nov\,08 & \textsf{20081127.071936} &      &      \\
28\,Nov\,08 & \textsf{20081128.011935} &      &  \\
28\,Nov\,08 & \textsf{20081128.121937} &      &  \\
28\,Nov\,08 & \textsf{20081128.191937} &      &  \\
29\,Nov\,08 & \textsf{20081129.011934} &      &  \\
29\,Nov\,08 & \textsf{20081129.071936} &      &  \\
29\,Nov\,08 & \textsf{20081129.121938} &      &  \\
29\,Nov\,08 & \textsf{20081129.201937} &      &  \\
\hline
26\,Aug\,10 & \textsf{20100826.011940} & 26\,Aug\,10 & 01:00:09.13 \\
26\,Aug\,10 & \textsf{20100826.131938} &      &  \\
27\,Aug\,10 & \textsf{20100827.011941} &      &  \\
28\,Aug\,10 & \textsf{20100828.033621} &      &  \\
 \hline
 &                            & 20\,May\,11 & 01:00:08.12 \\
 &                            & 20\,May\,11 & 02:06:44.12 \\
 &                            & 20\,May\,11 & 03:13:20.12 \\
 &                            & 20\,May\,11 & 04:20:08.12 \\
 &                            & 20\,May\,11 & 05:26:44.12 \\
 &                            & 20\,May\,11 & 06:33:20.12 \\
\hline
22\,Jun\,12 & \textsf{20120622.011940} & 22\,Jun\,12 & 06:00:08.12 \\
22\,Jun\,12 & \textsf{20120622.131941} & 22\,Jun\,12 & 08:15:08.13 \\
23\,Jun\,12 & \textsf{20120623.131940} & 22\,Jun\,12 & 10:29:56.12 \\
 &                            & 22\,Jun\,12 & 12:44:56.12 \\
 &                            & 22\,Jun\,12 & 14:59:56.14 \\
\hline
13\,Sep\,13 & \textsf{20130913.011942} & 15\,Sep\,13 & 01:00:07.14 \\
13\,Sep\,13 & \textsf{20130913.131944} & 15\,Sep\,13 & 03:12:07.12 \\
14\,Sep\,13 & \textsf{20130914.131941} & 15\,Sep\,13 & 05:24:07.12 \\
15\,Sep\,13 & \textsf{20130915.011941} & 15\,Sep\,13 & 09:47:55.13 \\
15\,Sep\,13 & \textsf{20130915.131943} & 15\,Sep\,13 & 10:53:55.12 \\
16\,Sep\,13 & \textsf{20130916.011941} &      &  \\
\hline
16\,Jul\,14 & \textsf{20140716.011946} & 16\,Jul\,14 & 01:00:07.14 \\
16\,Jul\,14 & \textsf{20140716.131941} & 16\,Jul\,14 & 05:24:07.12 \\
 &                            & 16\,Jul\,14 & 07:35:55.13 \\
 &                            & 16\,Jul\,14 & 09:47:55.12 \\
 &                            & 16\,Jul\,14 & 11:59:55.12 \\
\hline
28\,May\,15 & \textsf{20150528.011944} & 28\,May\,15 & 01:15:07.12 \\
\hline
\end{tabular}
 \vspace{-0.03\textwidth}
\end{table}

Table 1 shows that some EIT and AIA dates match each other, \textit{e.g.} for 22 June 2012, 15 September 2013, 16 July 2014, and 28 May 2015, which increases the separation from the instrument-based issues in favor of solar-based results.

\section{Data Reduction}
\label{S-Data Reduction}

Data reduction for this analysis was updated compared to that described by
\inlinecite{Didkovsky14a}.
There are three reasons to update the previous data-reduction algorithm. First, a longer time interval (1996\,--\,2015) has been analyzed in this work compared to 1996\,--\,2011 in the previous analysis. This longer time interval which includes some dates outside the solar minima of 1996 and 2008\,--\,09, requires better filters and some additional procedures to clean the data. Second, starting with 2010 when SDO/AIA images became available, the current analysis compares pairs of SOHO/EIT and SDO/AIA spatial spectra. Such a comparison benefits from the points of view of different responsivity and degradation for the two instruments with significantly different times of operation in space. However, they have different pixel scales, 2.6 and 0.6~arcsec per pixel for EIT and AIA, respectively. Thus, the spatial frequency units pix$^{-1}$ used in
\inlinecite{Didkovsky14a}
are changed to Mm$^{-1}$ in this work. Third, the initial step in determination of spectral ratios in the previous analysis was to match the spectral densities at the highest spatial frequencies in vicinity of 0.5~pix$^{-1}$ by multiplying more recent spectra by a coefficient which compensates both solar and instrumental, \textit{e.g.} degradation, sources of spectral density differences. Such a method of determination of spectral ratios for two pairs of spectra from successive time intervals 1996 \textit{vs.} 2010 and 2010 \textit{vs.} 2015 is shown in Figure 1.

   \begin{figure}    

   \centerline{\hspace*{0.015\textwidth}
               \includegraphics[width=0.51\textwidth,clip=]{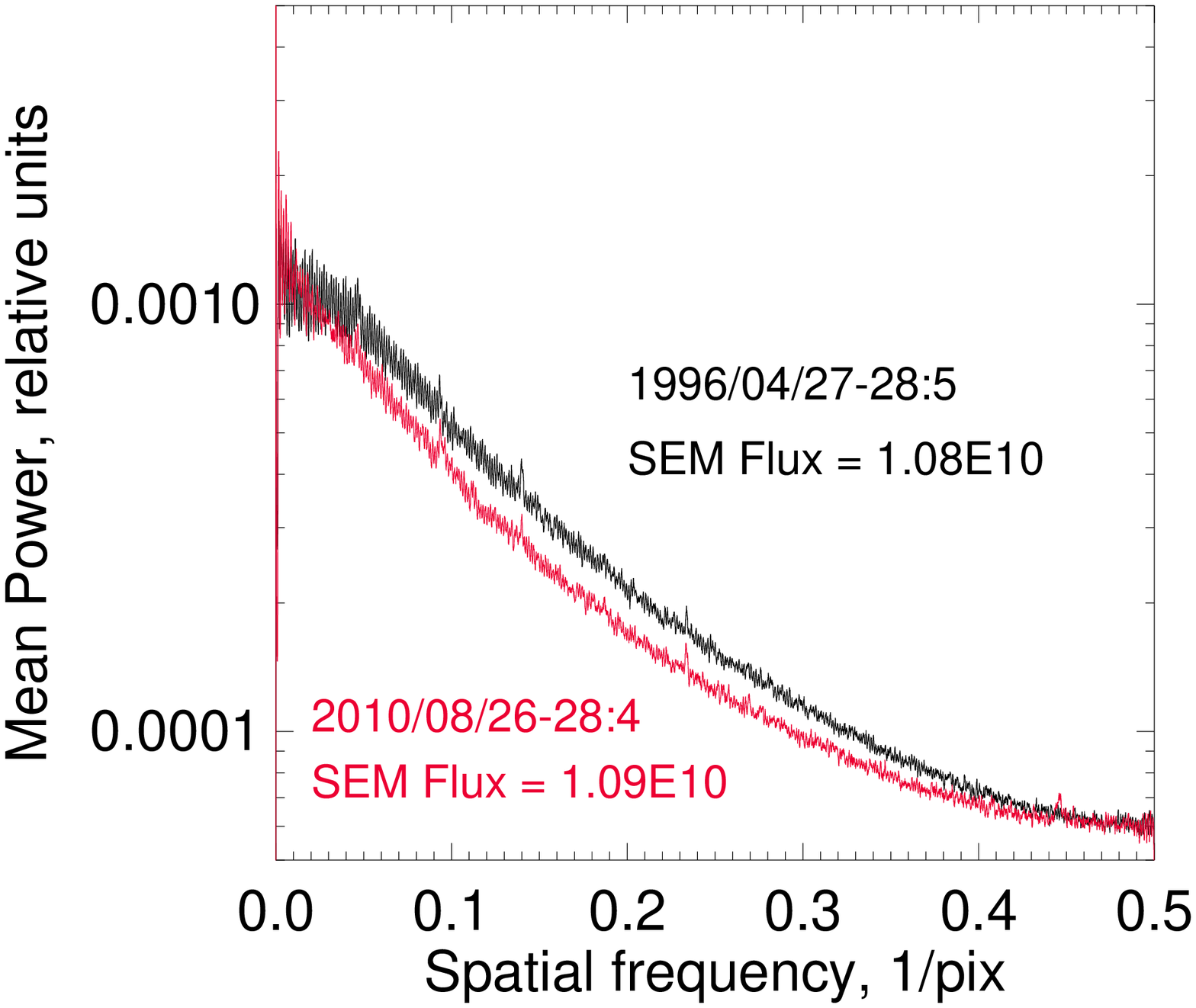}
               \hspace*{-0.03\textwidth}
               \includegraphics[width=0.51\textwidth,clip=]{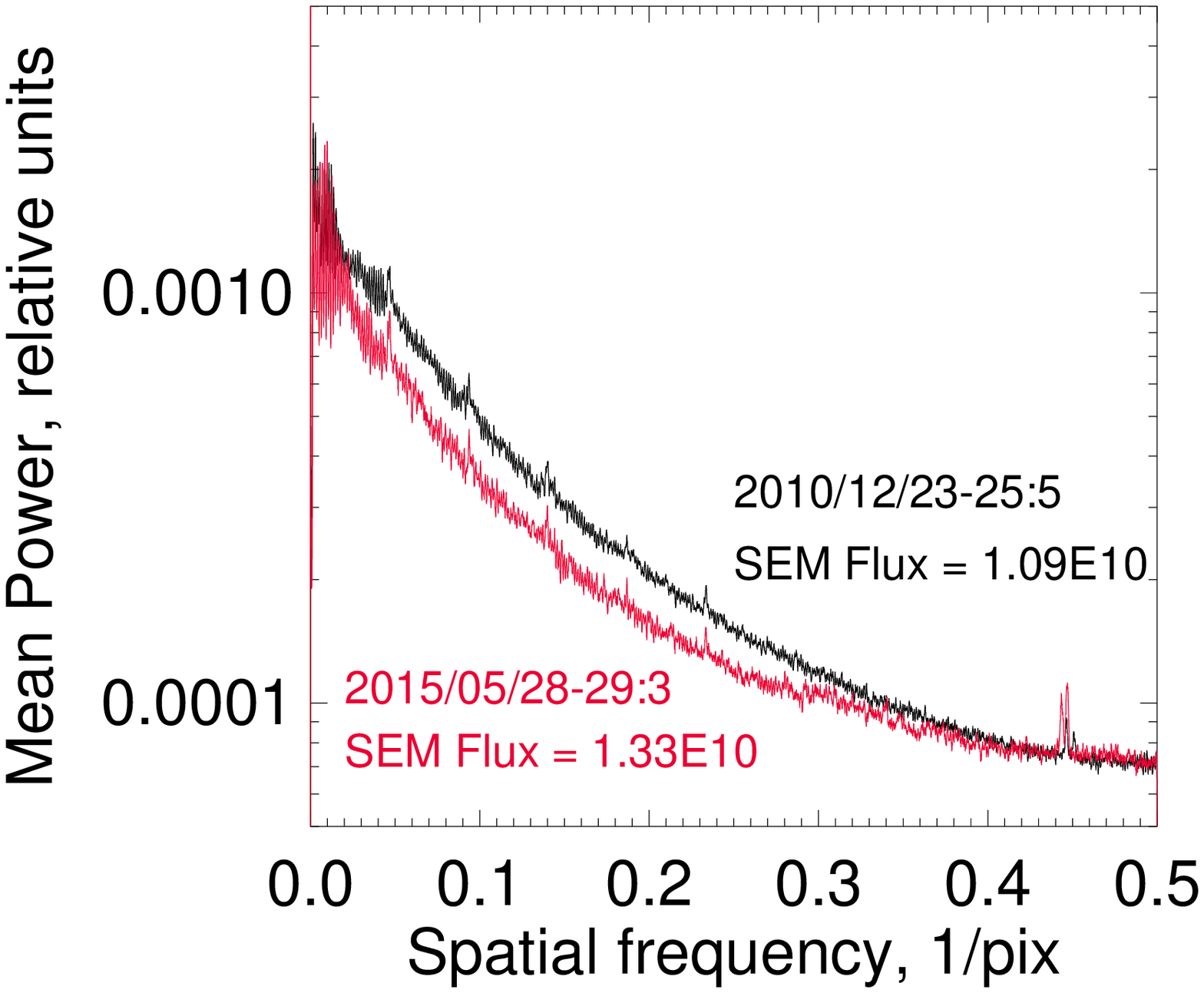}
              }
     \vspace{-0.35\textwidth}   
     \centerline{\Large \bf     
      \hspace{0.0 \textwidth}  \color{white}{(a)}
      \hspace{0.415\textwidth}  \color{white}{(b)}
         \hfill}
     \vspace{0.31\textwidth}    

\caption{Left: a comparison of EIT power spectra for 1996 (black) and 2010 (red). Right: the same for 2010 (black) and 2015 (red). These two pairs of spectra show continuously decreased power for two consecutive time intervals, from 1996 to 2010 and from 2010 to 2015. The spatial power at the mid-spatial frequencies are smaller for the 2010 spectra compared to the 1996 spectra (left) with about the same SEM first-order daily-mean fluxes, 1.09$\times$10$^{10}$ and 1.08$\times$10$^{10}$ ph cm$^{-2}$ s$^{-1}$, correspondingly.
        }
   \label{Fig1}
   \end{figure}

Calculated spectral density arrays were replaced with \textsf{median} (an IDL procedure) filter curves using 355 and 1501 integration data points for EIT and AIA spectra, correspondingly. Each of the plots shown in Figure 1 is a result of the averaged filter curves for five, four, or three individual spectra for 1996, 2010, and 2015, respectively. This approach was used by
\inlinecite{Didkovsky14a}
to compare mid-frequency spatial power is based on comparing adjusted spectra (more recent were multiplied by a coefficient) to match the power of the reference, \textit{e.g.} 1996 spectra at the highest spatial frequency in the vicinity of 0.5~pix$^{-1}$.
With significantly different pixel scales for EIT and AIA, determination of the spectral ratios is now based on the matching highest ``solar'' spatial frequencies, which are different for EIT and AIA. In addition to these three updates, a new cleaning procedure to limit large-amplitude peaks related to sharp (one or two pixels) increases from the energy deposited by energetic particles has been used for analyzing a longer (1996\,--\,2015) time interval. The details of the new approaches and tests are described in Section 4 below.

\section{Previous Tests and New Approaches}
\label{S-Previous Tests and New Approaches}

\subsection{Previous Tests}

The conclusion that the results of solar disk spatial power spectra ratios reported by
\inlinecite{Didkovsky14a}
reflected some solar changes was based on five tests to verify that the changes were not related to the instrumental sources such as degradation and changes of dark and light pixel responses. One of the tests showed that restoring of decreased spatial power at mid-size frequencies for more recent dates to match the power in the reference spectrum (\textit{e.g.} for 1996) may be provided by artificially stretching a more recent image, thus increasing the dimensions of the observed EUV segmentation cells and compensating for temporal changes of the network structures. Another finding was that some more recent (\textit{e.g.} 2010 \textit{vs.} 1996, see Figure 1, left) spectral-ratio decreases occurred for exactly the same absolute solar EUV flux measured by the SEM 30.4~nm channels as in 1996. For the times analyzed with increased solar activity and larger SEM fluxes, the ratios continued to decrease indicating no detectable dependence from the SC activity variations.

\subsection{Removing Pixel Signals Related to Energetic Particles}

In contrast to the procedure of removing energetic particle signals by setting up a fixed threshold used in
\inlinecite{Didkovsky14a},
the limiting thresholds for this analysis were implemented as follows. First, a 1D array derived from the 2D on-disk image effective counts (dark images were subtracted and pixel non-uniformity corrected using lamp images) was filtered with two curves: one filter $F_{1}(k)$ with a smoothing window [$w1$] of five pixels, another filter $F_{2}(k)$ with a window [$w2$] of 11 pixels. Then for the $k$-th pixel of the 1D array the resulting signal $C_{\rm eff} (k)$ was equal to either a filter signal [$F_{2}(k)$, see Equation (1)] or a corresponding image signal, $C_{\rm eff} (i,j)$ [Equation (2)]:

 \begin{equation}       
C_{\rm eff}(k)=F_{2}(k)\quad    \mathrm{if} \quad    C_{\rm eff}(i,j) > F_{1}(k)*\alpha
\end{equation}
 \begin{equation}       
C_{\rm eff}(k)=C_{\rm eff}(i,j)\quad    \mathrm{if} \quad    C_{\rm eff}(i,j) \leq F_{1}(k)*\alpha
\end{equation}
where the $\alpha$-coefficient was determined as 1.5 by analyzing the effect of filtering. A lower number than 1.5 would lead to the change of solar-image segmentation-cell structure by replacing a number of observed pixel signals with a lower contrast combination of pixel signals from the filtered array using $F_{2}(k)$ data. A higher number decreases the number of removed mid-amplitude sharp energetic-particle peaks and would produce the results similar to the filters with fixed thresholds. Thus, the use of window [$w1$] with five-pixel window for the filter $F_{1}(k)$ and the $\alpha$=1.5 allowed us to distinguish between the sharp increases related to solar intensity and those related to energetic particles. Our analysis of the 1D arrays showed that about 95\,\% of the energetic-particle peaks were replaced with the mean value from the [$w2$]-window of 11 pixels using Equation (1). For most of the solar-image pixels not affected by the energetic particle signals the data for the 1D array were consistent with the data reduced with the fixed thresholds. This filtering effect is shown in Figure 2, (right) as compared to the unfiltered data in Figure 2, (left) for the same 1D array from the image \textsf{efz20120622.131941}.
   \begin{figure}    

   \centerline{\hspace*{0.015\textwidth}
               \includegraphics[width=0.515\textwidth,clip=]{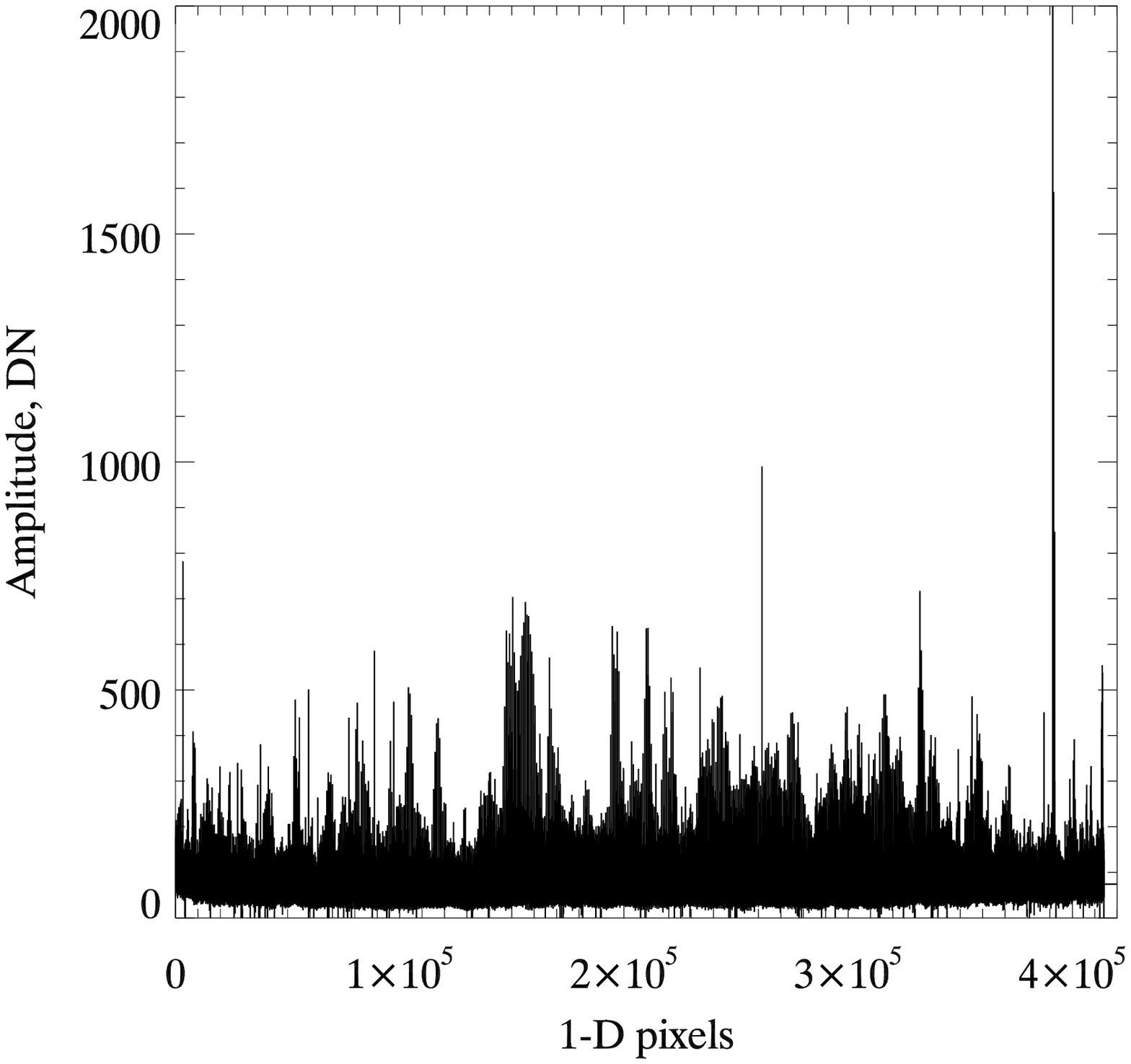}
               \hspace*{-0.03\textwidth}
               \includegraphics[width=0.515\textwidth,clip=]{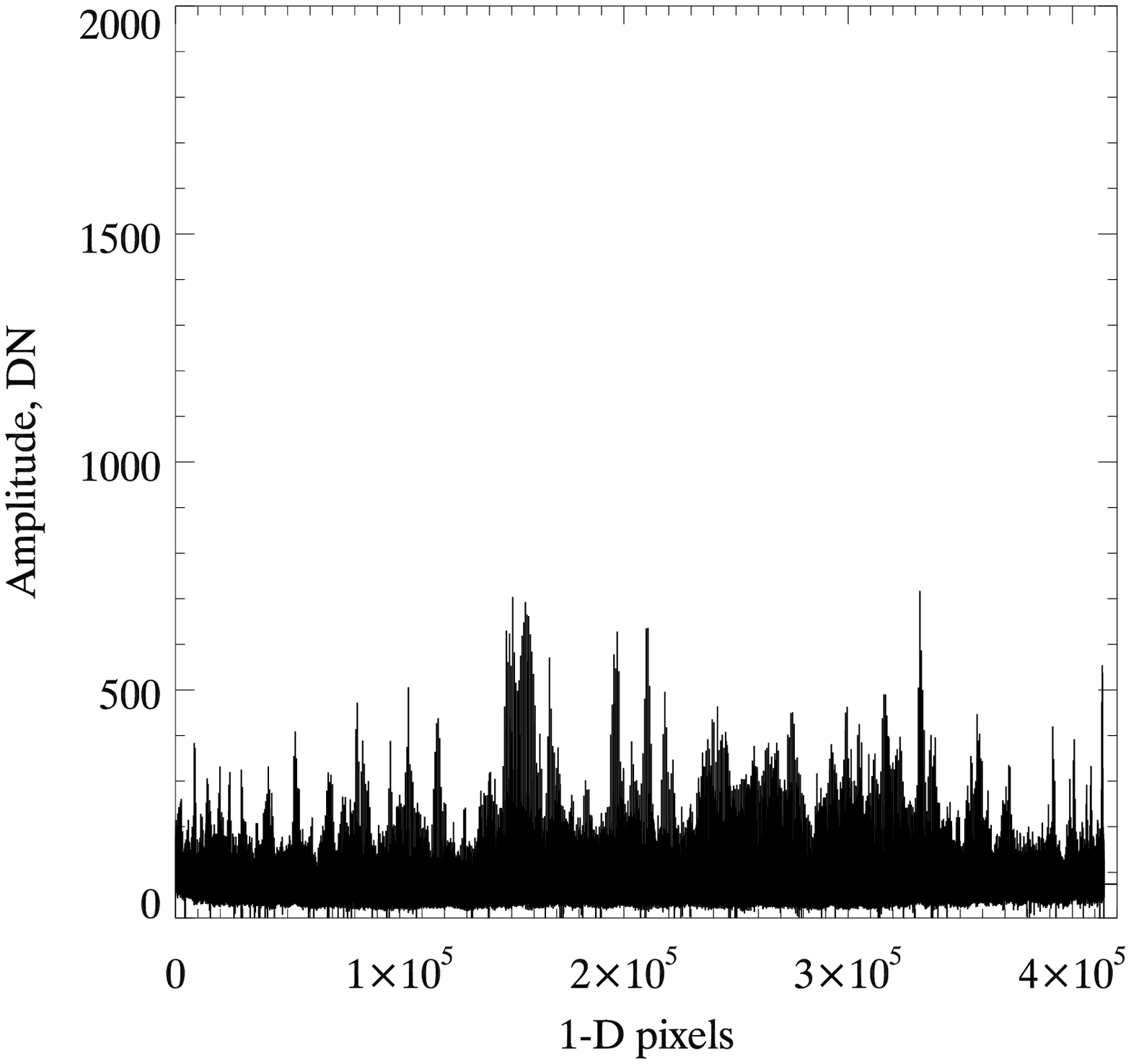}
              }
     \vspace{-0.35\textwidth}   
     \centerline{\Large \bf     
      \hspace{0.0 \textwidth}  \color{white}{(a)}
      \hspace{0.415\textwidth}  \color{white}{(b)}
         \hfill}
     \vspace{0.31\textwidth}    

\caption{Left: Unfiltered 1D array with a number of narrow energetic particle-related peaks, see, for example, peaks around 2.6 and 3.9$\times 10^{5}$ pixels; Right: the same but filtered using Equations (1) and (2). Note, that the chosen procedure preserves the solar features while it removes particle-related peaks.
        }
   \label{Fig2}
   \end{figure}

Finally, the remaining portion of the 1D array between the number of pixels from the 2D image and up to 2$^{19}$ pixels for EIT images (2$^{n}$ is used for the fast Fourier transform) was filled with a constant value, which was equal to the mean number of the effective counts from the solar disk.

\subsection{A Model to Compare Spectra Obtained from EIT and AIA with Different Spatial Resolution}

Solar EUV images observed by SOHO/EIT and SDO/AIA have different pixel scales: about 2.6 and 0.6~arcsec pix$^{-1}$, correspondingly. This different resolution affects the visibility of small-scale solar segmentation on the disk and, thus, the spectral density in the power spectra. The largest possible spatial frequency, a Nyquist frequency equal to a half of the sampling rate, which is 0.5~pix$^{-1}$, represents different solar scales on the solar-disk images observed by EIT and AIA. In contrast to our previous analysis in
\inlinecite{Didkovsky14a},
based mainly on power-spectra ratios calculated from the EIT images with the pix$^{-1}$ units of spatial frequency, this study is to use solar spatial frequency with units Mm$^{-1}$, which allows us to directly compare EIT and AIA spectra at the EIT-based spatial frequency. To model different EIT and AIA spatial frequency as functions of sizes of solar features, Equation (3) has been used:
 \begin{equation}       
\mu_{\rm i}= 1/(b_{i} \beta \Delta)
\end{equation}
where $\mu$ is the spatial frequency, $i$ ranges from 0 to 999; $\beta$ is the EIT or AIA pixel angular scale, 2.6 or 0.6~arcsec, correspondingly; $\Delta$ = 0.7~Mm arcsec$^{-1}$ is the solar angular scale, $b_{i}$ is the number of pixels required to cover a segment on the EIT or AIA image

\begin{equation}       
b_{i} = 1 + ceil(\alpha_{i}/ \beta )
\end{equation}
where the function `ceil' represents the closest integer greater than or equal to its argument, and $\alpha_{i}$
is the test array of angular segment sizes from 2.0~arcsec with 0.1~arcsec increment.

The initial angular size of 2.0~arcsec that may be covered by a maximum of two EIT pixels sets an upper limit on the spatial frequency. This limit is equal to the previously used spatial frequency of 0.5 pix$^{-1}$ converted to the Mm$^{-1}$ units: 0.27 Mm$^{-1}$. Figure 3 shows the modeled spatial frequency as a function of inverted segment sizes for EIT and AIA instruments. The inverted sizes specified on the horizontal axis in Figure 3 are $1/(\alpha_{i} \Delta)$. Assuming image motion during the exposure, \textit{e.g.} such as solar rotation and a jitter, the image of the smallest segment with about 1.4~Mm (see the right-hand points in Figure 3) is covered by two EIT pixels, or five AIA pixels. For example, $\mu_{EIT}$ ($\mu_{AIA}$) based on Equation (3) for two EIT (five AIA) pixels are 0.27 Mm$^{-1}$ for EIT and 0.48 Mm$^{-1}$ for AIA. Note, these frequencies remain unchanged (have the same ordinates) for a number of test inverted sizes until the inverted size increments are within initial number of pixels, two for EIT or five for AIA. Because EIT has much lower angular resolution than AIAD, the number of horizontal points for EIT spatial frequency (stars) is larger compared to the AIA frequency (diamonds). For example, EIT spatial frequency of 0.18~ Mm$^{-1}$ corresponds to the range of segment sizes from about 3.7 to 5.5~Mm.
\begin{figure}    
   \centerline{\includegraphics[width=0.95\textwidth,clip=,
                  bb=24 20 760 350]{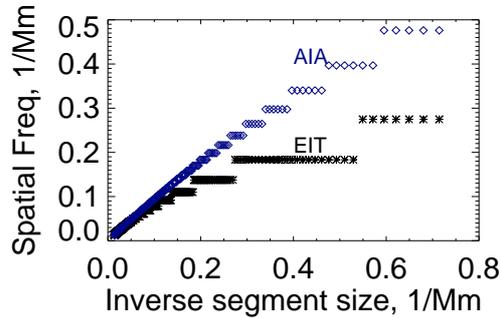}
                 }
                            \vspace{0.05\textwidth}   
              \caption{Modeled EIT (stars) and AIA (diamonds) spatial frequencies [Mm$^{-1}$] as a function of the inverse segment size using the largest possible number of pixels to cover the image of the segment.
                      }
   \label{Fig3}
   \end{figure}
The model (Figure 3) shows that EIT and AIA spatial-frequency values start to be different at and below about 10~ Mm structures and reach 0.27 and 0.48 Mm$^{-1}$ spatial frequency for EIT and AIA, correspondingly, for small-size segments. These modeled $\mu_{\mathrm EIT}$ and $\mu_{\mathrm AIA}$ values are used to compare EIT and AIA spectral ratios at the highest spatial frequencies, for which we expect the greatest sensitivity in the spatial spectra to reflect the solar changes. However, due to a step-like character of the spatial frequency related to the increment of affected EIT pixels with a small increase of the segment size, \textit{e.g.} from 2.6 (may be covered by two pixels) to 2.61~arcsec (it requires three pixels), a comparison at the spatial frequencies one step below the edge frequencies could be useful. These frequencies are 0.18 and 0.40 Mm$^{-1}$ for EIT and AIA, correspondingly.

\subsection{Spatial Power Spectra}

In addition to the tests described in Subsection 4.1 and presented by
\inlinecite{Didkovsky14a},
some new tests were performed to extend the analyses to a more confident level. These tests include a comparison of the spectra from the whole solar disk and from the central portion ($R$ = 0.7\,R$_{\odot}$), a verification of the model (Figure 3) using a comparison of AIA and EIT spectra, a comparison of the spectra based on the use of columns (instead of rows) to form the 1D array, and a comparison of the spectra with added pixel random noise to the 1D array.

Figure 4 compares the spatial power spectra for AIA images obtained on 28 May 2015 and 26 August 2010 with the spectra derived from the whole solar disk (left) and the central portion of the disk with $R$ = 0.7\,R$_{\odot}$ (right).
   \begin{figure}    

   \centerline{\hspace*{0.015\textwidth}
               \includegraphics[width=0.55\textwidth,clip=]{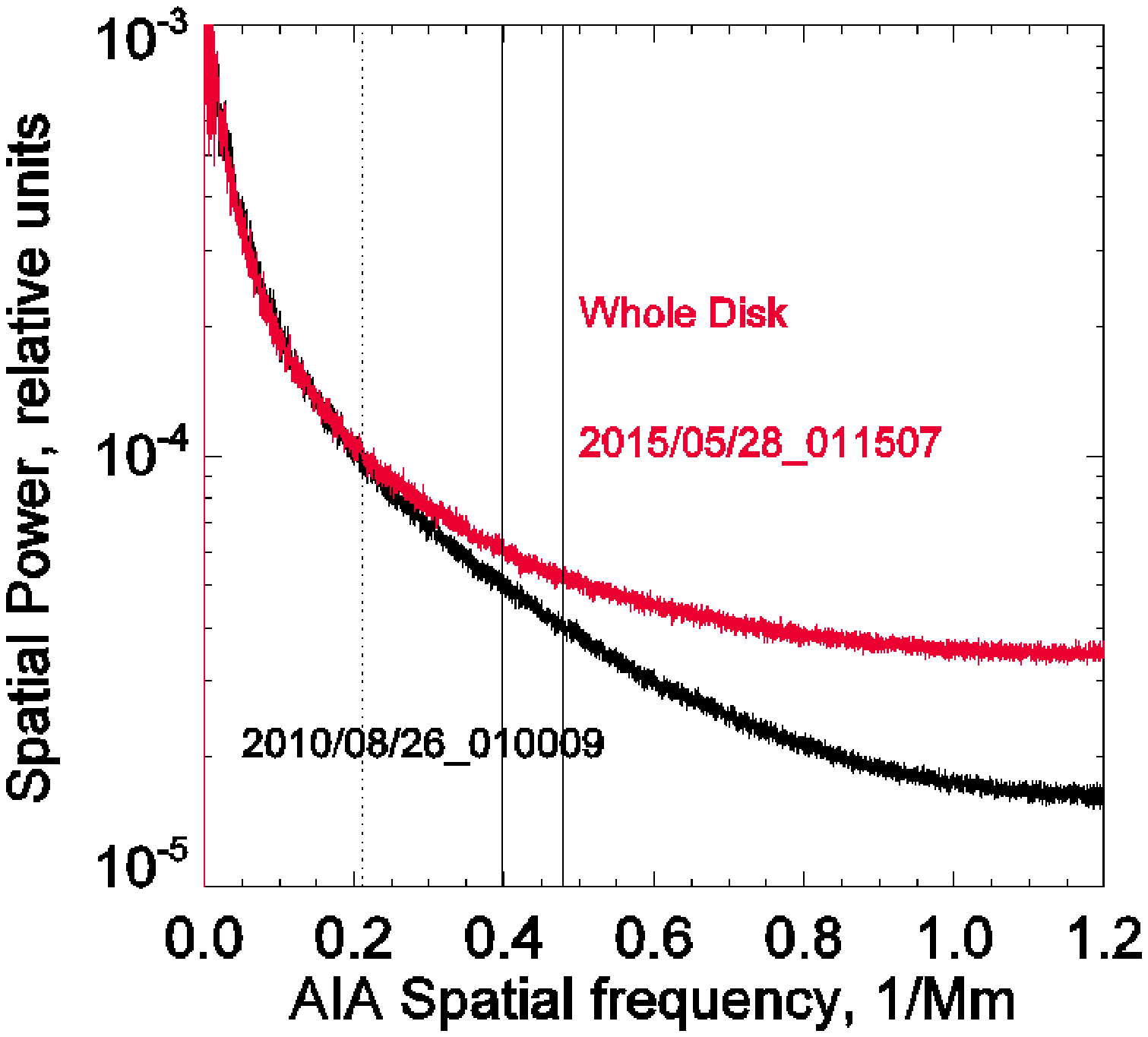}
               \hspace*{-0.03\textwidth}
               \includegraphics[width=0.55\textwidth,clip=]{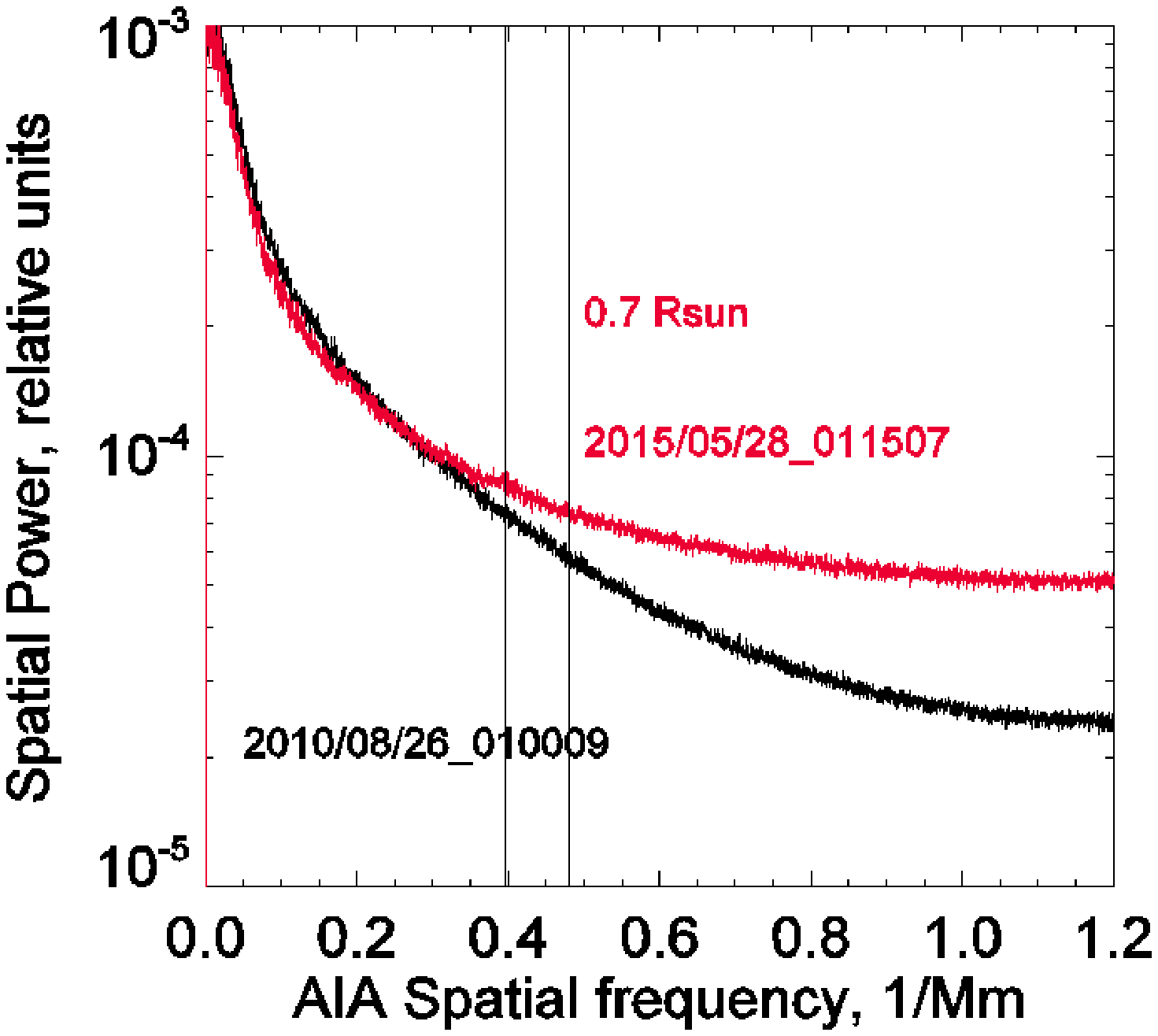}
              }
     \vspace{-0.35\textwidth}   
     \centerline{\Large \bf     
      \hspace{0.0 \textwidth}  \color{white}{(a)}
      \hspace{0.415\textwidth}  \color{white}{(b)}
         \hfill}
     \vspace{0.11\textwidth}    

\caption{A comparison of spatial power spectra for 28 August 2010 (black) and 28 May 2015 (red) for the whole solar disk (left) and a central portion of the disk with 0.7\,R$_{\odot}$ (right). Vertical solid lines correspond to the modeled spatial frequencies of 0.397 and 0.48\,Mm$^{-1}$ which provide a comparison for the spectral ratios between AIA and EIT. Dotted line (left) shows spatial frequency (0.21\,Mm$^{-1}$) at which the black and red curves have the same spectral density.
        }
   \label{Fig4}
   \end{figure}
In contrast to the pairs of spectra shown in Figure 1, the spectra in Figure 4 were plotted ``as they are”, without matching the curves at the highest spatial frequency. Without such matching, the newer spectrum shows some increased power at high spatial frequency, which is consistent with the decreased power in the newer spectrum at the mid-spatial frequencies for the case of matching the curves. The power-spectra ratios for 2015 (red) to 2010 (black) at the spatial frequency of 0.397\,Mm$^{-1}$ (left vertical lines) are 1.22 and 1.20 for the whole disk (Figure 4, left) and the central portion (Figure 4, right), correspondingly. The spectral-density-ratios for the right vertical lines at the spatial frequency of 0.48\,Mm$^{-1}$ are 1.33 and 1.32. The vertical lines shown in Figure 4 correspond to the highest modeled spatial frequencies shown for AIA in Figure 3 (top right diamond points).This test shows a small difference for the spectra ratios using the two approaches (whole disk and a central portion).
	For the same pair of dates as shown in Figure 4, power spectra based on EIT data (Figure 5, left) show about the same ratio of 1.27. Figure 5, (right) shows the same spectra but adjusted to match other at the highest spatial frequency using the approach described in
\inlinecite{Didkovsky14a}.
\begin{figure}    
  \centerline{\hspace*{0.015\textwidth}
               \includegraphics[width=0.515\textwidth,clip=]{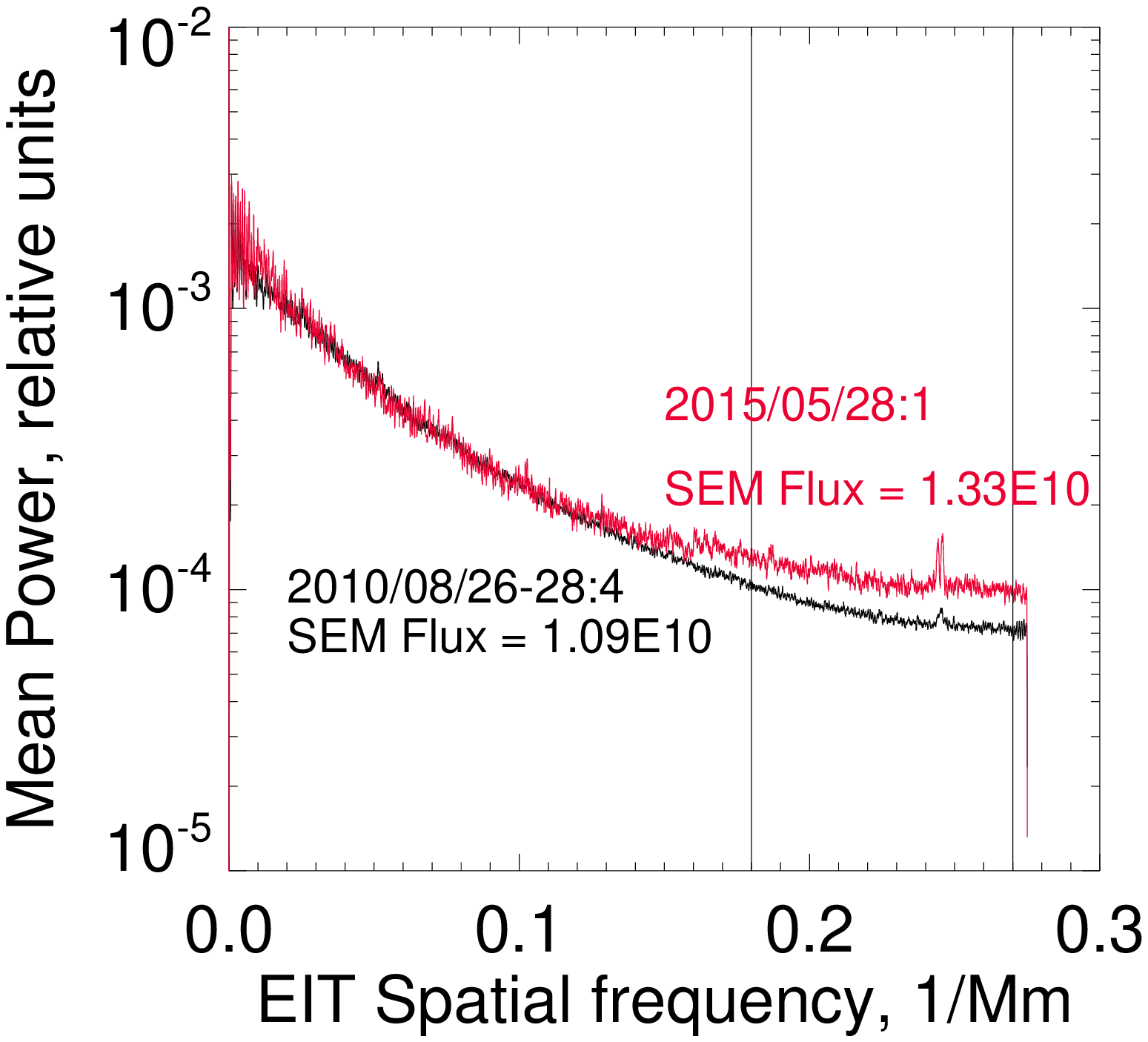}
               \hspace*{-0.03\textwidth}
               \includegraphics[width=0.515\textwidth,clip=]{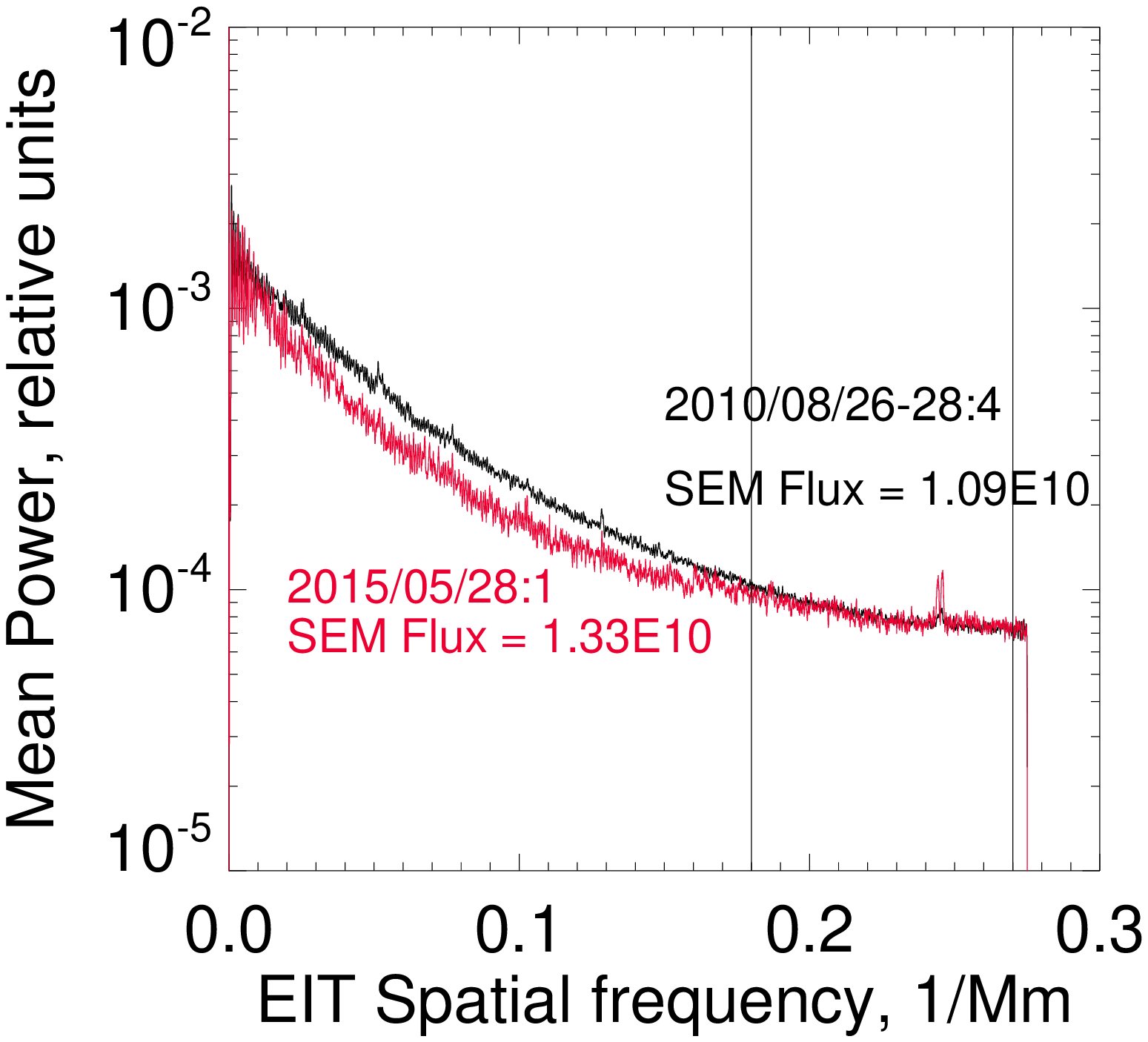}
              }
     \vspace{-0.35\textwidth}   
     \centerline{\Large \bf     
         \hfill}
     \vspace{0.31\textwidth}    
              \caption{Left. A comparison of EIT power spectra for 26\,--\,27 August 2010 (black) and 28 May 2015 (red). The vertical lines correspond to the spatial frequency of 0.18 Mm$^{-1}$ and 0.27 Mm$^{-1}$ determined from the model. The 2015/2010 ratio at the left frequency axis point is 1.27, and at the right frequency axis point is 1.35. This numbers are close (within 4\,\%) to the pair 1.22 and 1.33 from the AIA spectra shown in Figure 4. Right. The same spectra as in the left panel but with matching the spectral densities at the highest spatial frequency of 0.27\,Mm$^{-1}$ which is close to the spatial frequency of 0.5~pix$^{-1}$ as reported by
\inlinecite{Didkovsky14a}.
                      }
   \label{Fig5}
   \end{figure}

Figure 5, (left) shows that the spectral density increase starts at EIT spatial frequency of about 0.15~Mm$^{-1}$ which corresponds to segment sizes from about 4.9 to 5.5~Mm based on Equations (3) and (4). The uncertainty is related to the EIT limited angular resolution. The right panel shows that a decreased section of the spectrum starts at about 0.17~Mm$^{-1}$, which corresponds to the same range (4.9 to 5.5~Mm) of segment sizes.
The options shown in Figure 5 with the spectra as they were calculated (left) and the spectra (right) with a match of the spectral densities at the highest spatial frequency may be used to estimate levels of temporal change in the relative spectral densities for each. The first option for the AIA spectra is shown in Figure 4, (left). This option demonstrates some temporal increase (red line) of the small-size segments toward the highest spatial frequency of 0.48~Mm$^{-1}$ (right solid-vertical line). Another option (not shown) is based on the match of the densities at the AIA highest spatial frequency. This second option makes the zero difference between the red and black curves at this point and demonstrates a decrease of the density for the newer (2015) spectrum (red line) toward lower spatial frequencies. This estimation of the total temporal changes of densities $\Delta A$ and $\Delta B$ for the spectra $a0$ and and $b0$ (black and red curves in Figure 4, (left), respectively) is given by Equations (5) and (6):

\begin{equation}       
\Delta A = \sum_{f1}^{f2}(a0 - b0)
\end{equation}
\begin{equation}       
\Delta B = \sum_{f1}^{f2}(b0 - a0)
\end{equation}

 We have chosen a frequency range from $f$1 to $f$2 based on Figure 4, (left): $f$1 = 0.21~Mm$^{-1}$ at the dotted vertical line, for which the spectral densities from the red and black curves are equal and $f$2 = 0.48~Mm$^{-1}$, which corresponds to the right edge of the AIA spatial frequency. $\Delta A$ and $\Delta B$ for this chosen range of the frequencies are 8.4 and 8.5, correspondingly. Comparing Figures 4 (AIA) and 5 (EIT) we found that the AIA-measured segment size for the point where the densities start to increases $f$1 = 0.21~Mm$^{-1}$ is from 4.3 to 4.6~Mm. This range is close to the one (4.9 to 5.5~Mm) determined from the EIT (Figure 5) spectra but has a smaller uncertainty.

\subsection{A Test with a Replacement of Solar Disk Image Rows by Columns}

The method of forming a 1D array from the 2D CCD solar-disk image used by
\inlinecite{Didkovsky14a}
was to use CCD rows. A test forming a 1D array using columns instead of rows does not change any results from the power spectra.

\subsection{A Test with Added Random Pixel Read-Out Noise}

Some increase in power density with time at higher spatial frequencies could be related to increasing time-dependent CCD pixel random noise. To assess the sensitivity of power spectra to increased random noise in EIT (a longer operation time compared to the AIA), noise was artificially added in the form of randomly applied offsets to the pixel DN values at various levels to the EIT spatial 1D data sequence for 18 May 2010 and the ``noisy” power spectra were compared to both the original 2010 power spectrum and the power spectrum from 29 May 2015 (Figure 6).
One source of such noise is dark-current noise, which depends on the CCD temperature affected by orbital position of the Earth around the Sun. In addition to some sensitivity of the dark current and its RMS noise to the temperature change, another source of time-dependent random noise is read-out noise with its on-chip and off-chip components (\textit{e.g.} from off-chip analog to data converter which could become more noisy due to the increased amount of the total ionizing dose). The comparison (Figure 6) shows that the addition of pixel noise at a level of approximately 6.4 DN RMS results in an increase in spectral power for higher spatial frequencies that is comparable to the increase observed between the 2010 and 2015 dates considered in this study. It is expected that if this 2010-to-2015 increase is related to increased readout noise, then the noise increase should also be evident in comparisons of RMS for EIT dark images between these two dates. However, no such increase in noise is evident in the dark images – spatial variability in pixel dark values remains at about three DN RMS for dark images through the time interval from 2010 to 2015 with different months and orbit-related temperatures, ranging from May (2010, 2011, and 2015) to October (2014), thus suggesting that the increase in spectral power at higher spatial frequencies is the result of changes in solar EUV network structure and not related to an increase in detector readout noise.

\begin{figure}    
   \centerline{\includegraphics[width=1.15\textwidth,clip=,
                  bb=4 420 760 750]{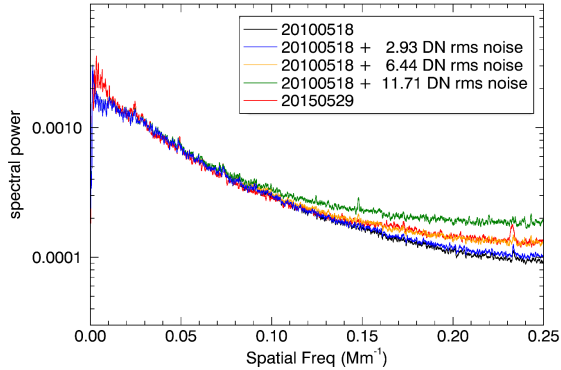}
                 }
                            \vspace{-0.05\textwidth}   
              \caption{Comparison of EIT power spectra from 18 May 2010 (black curve) and 29 May 2015 (red curve overlapped with the orange curve) with 2010 spectra to which noise at various RMS levels has been artificially added (blue, orange, and green curves). It can be seen that the increase in spectral power observed at high-spatial-frequencies between 2010 and 2015 would require an increase in pixel readout noise of approximately 6.4\,DN RMS. Spatial variability of pixel signal in EIT dark images however remains virtually constant at around 3\,DN RMS throughout the 2010 to 2015 time interval suggesting that the increase in high-spatial-frequency spectral power observed in EIT solar images between 2010 and 2015 is related to changes in solar EUV network structure and not to increased detector noise.
                      }
   \label{Fig6}
   \end{figure}

The test with adding to the 1D pixel signal array a random noise shows that such noise at the level of 3 DN RMS determined for the EIT dark-image data does not add significantly to the original (noise-less) spectrum\,--\,compare black and red curves in Figure 6.

\section{Results}

The results of this work are spatial-power ratios for each of the years analyzed (see Table 1). The ratios were determined using April 1996 spectra as reference for EIT and May 2010 as reference to AIA. Figure 7 shows EIT and AIA ratios at the matching spatial frequencies of 0.18 and 0.40\,Mm$^{-1}$ (one step to the left from the edge frequencies in Figure 3).
\begin{figure}    
   \centerline{\includegraphics[width=1.1\textwidth,clip=,
                  bb=4 420 760 750]{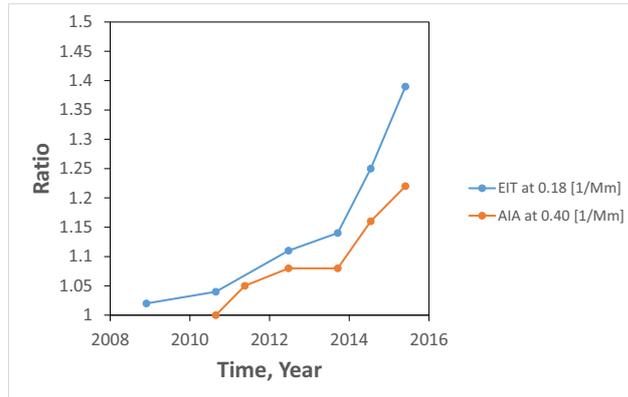}
                 }
                            \vspace{0.05\textwidth}   
              \caption{The changes of the ratios 1996/Year for EIT spectra (blue) and 2010/Year for the AIA spectra (orange).
                      }
   \label{Fig7}
   \end{figure}

Figure 7 shows a similar increase of the spectral ratios for the two EUV instruments, EIT and AIA with different mission time for the AIA reference point in 2010: EIT has been operating for 15 years and AIA for a couple of months. The ratios are compared to each other using matching spatial frequencies determined from the model shown in Figure 3 using 0.18 and 0.40\,Mm$^{-1}$ frequencies left vertical lines of Figures 5 and 4 for EIT and AIA, respectively. These frequencies correspond to detection of EUV network segmentation-cell-structure changes with a typical sizes of the fragments around two Mm.

\section{Conclusions}

This article advances the investigation presented by
\inlinecite{Didkovsky14a}
in three directions. The first is an extension of the analyzed time interval, from 1996 to 2015 compared to the previously analyzed, from 1996 to 2011, which spans more than 1.5 solar cycle. The second is a switch from the pixel-oriented spatial frequency which is different for EIT and AIA, to the solar-oriented spatial frequency. Based on the solar-oriented approach and modeled instrumental frequencies (Figure 3), the third advancement is a number of direct comparisons of the spatial spectra between EIT and AIA after AIA data became available in 2010. Results of the temporal \textit{increases} of spectral densities of small-size solar EUV segments (ranges from about 4.3 to 4.6~Mm) presented in this article are consistent with the range of segment sizes which are associated with the spectral density \textit{decreases}. Figures 4 and 5 and the analysis of the plots shown on these figures demonstrate this consistency. Figure 5 allowed us to explain two interpretations of the same physical process as a transformation (dissipation) of larger segments to smaller segments.

In addition to previously tested instrumental sources of the changes of spatial power, this work extended such tests to include important new tests and approaches to remove the spikes associated with energetic particles, to test two versions (rows or columns) for forming the 1D data arrays, a test with the use of the whole solar disk or a central portion with 0.7~R$_{\odot}$, and a test with adding to the 1D array data a random noise. Neither of these additional tests suggested an instrumental source of the ratio changes. Finally, the previous method of expressing the spatial frequency in pix$^{-1}$ units was extended to the use of ``solar” units [Mm$^{-1}$], which allowed us to compare results from two instruments, EIT and AIA for a much longer time interval than reported by
\inlinecite{Didkovsky14a}.
Due to different pixel scales for EIT and AIA, a model of their responses to different sizes of EUV segments has been developed. The model provided spatial frequencies for the two instruments, which could be used for the comparison of spectra at high frequencies. In contrast to the approach of aligning the high-frequency edges of the spectra \citep{Didkovsky14a}, in this work the spectra were analyzed ``as they appeared”, without any change of the spectral density to match the high-frequency edges.

The spectral densities from each spectrum analyzed represent relative spatial power of the full-disk EUV solar segmentation network. The ratios (Figure 7) were obtained from the spatial power spectra of individual days averaged in each of the groups of images confined by the horizontal lines in Table 1. This article shows a temporal evolution of solar EUV segmentation network as a dissipation of some mid-size (5~Mm and above) segments into smaller-size segments (from 1.7 to about 4.5~Mm based on the AIA spectra). Although determination of other characteristics of the dissipation of the observed EUV segments in the transition region is outside of the scope of this article, we found that the increase of the spectral density ratios at the high spatial frequencies (small segments) during quiet Sun conditions continues after 1996 through the analyzed interval to 2015 and is beyond the timescale of typical solar-cycle behavior.

\begin{acks}
 This work was supported in part by NASA/University of Colorado subcontract 153-5979.

\medskip\noindent\textbf{Disclosure of Potential Conflicts of Interest} The authors declare that they have no conflicts of interest.

\end{acks}

\end{article}

\end{document}